\documentclass[aps,twocolumn,a4paper,floatfix,amsfonts]{revtex4-1}                    

\usepackage{microtype}
\usepackage{graphicx}
\usepackage{wrapfig}
\usepackage{amsmath,amssymb,amsfonts}
\usepackage{mathtools}
\usepackage{bm}
\usepackage{color}
\usepackage[english]{babel}
\usepackage{time}

\newcommand{\tsub}[1]{\ensuremath{_{\mathrm{#1}}}} 

\newcommand{\kb}{k\tsub{B}}

\newcommand{\ie}{{\textit{i.\,e.\@}}}
\newcommand{\eg}{{\textit{e.\,g.\@}}}
\newcommand{\cf}{{\textit{cf.\@}}}

\newcommand{\todo}[1]{{\color{red}\textbf{TODO:}\,#1}}

\newcommand{\pspace}{\ensuremath{\Gamma} } 
\newcommand{\ospace}{\ensuremath{\Omega} } 
\newcommand{\cell}{\ensuremath{\mathcal{C}} } 
\newcommand{\vcell}{\ensuremath{\mu} } 

\newcommand{\partmap}{\ensuremath{ M} } 

\newcommand{\reals}{\ensuremath{\mathbb{R}}} 
\newcommand{\naturals}{\ensuremath{\mathbb{N}}} 

\newcommand{\abs}[1]{\ensuremath{\left\vert #1 \right\vert} }

\newcommand{\dyn}{\ensuremath{\Phi}}
\newcommand{\ja}{\ensuremath{\mathrm{J}}}

\newcommand{\Df}{\mathrm{D}}

\newcommand{\df}[1]{\ensuremath{\,\mathrm{d}#1\,}}

\renewcommand{\bar}{\ensuremath{\overline}} 

\newcommand{\set}[1]{{\ensuremath{\left\{#1\right\}}}}

\newcommand{\traj}[1]{\ensuremath{\underline{#1}} } 
\newcommand{\ifam}[1]{\ensuremath{\left(#1\right)}} 
\newcommand{\pden}{\ensuremath{\varrho}} 
\newcommand{\tind}[1]{^{(#1)}} 
\newcommand{\tprob}[2]{\ensuremath{a_{#2,#1}}} 

\newcommand{\fgden}{\ensuremath{\pden\tsub{fg}}}
\newcommand{\cgden}{\ensuremath{\pden\tsub{cg}}}
\newcommand{\priorden}{\ensuremath{{\pden}^*}} 

\newcommand{\entrv}{\ensuremath{s}}
\newcommand{\eprv}{\ensuremath{\Delta{\entrv}}}

\newcommand{\obsentrv}{\ensuremath{\entrv_{\mathrm{obs}}}} 
\newcommand{\priorentrv}{\ensuremath{\entrv^*}} 
\newcommand{\crossentrv}{\ensuremath{\entrv\tsub{cross}}} 
\newcommand{\contentrv}{\ensuremath{\entrv\tsub{cont}}} 

\newcommand{\relentrv}{\ensuremath{\entrv_{\mathrm{rel}}}} 

\newcommand{\sysentrv}{\ensuremath{\entrv_{\mathrm{sys}}}} 
\newcommand{\medentrv}{\ensuremath{\entrv_{\mathrm{med}}}} 
\newcommand{\totentrv}{\ensuremath{\entrv_{\mathrm{tot}}}} 

\newcommand{\sysent}{\ensuremath{\ent_{\mathrm{sys}}}} 
\newcommand{\medent}{\ensuremath{\ent_{\mathrm{med}}}} 
\newcommand{\totent}{\ensuremath{\ent_{\mathrm{tot}}}} 

\newcommand{\syseprv}{\ensuremath{\eprv_{\mathrm{sys}}}} 
\newcommand{\medeprv}{\ensuremath{\eprv_{\mathrm{med}}}} 
\newcommand{\toteprv}{\ensuremath{\eprv_{\mathrm{tot}}}} 

\newcommand{\entfun}{\ensuremath{\mathcal{H}}}
\newcommand{\dklfun}{\ensuremath{\mathcal{D}\tsub{KL}}}
\newcommand{\crossentfun}{\ensuremath{\mathcal{H}\tsub{cross}}}

\newcommand{\ent}{\ensuremath{S}} 

\newcommand{\fgent}{\ensuremath{\ent_{\mathrm{fg}}}} 
\newcommand{\cgent}{\ensuremath{\ent_{\mathrm{cg}}}} 
\newcommand{\relent}{\ensuremath{\ent_{\mathrm{rel}}}} 

\newcommand{\trav}[1]{\ensuremath{\langle\!\langle #1 \rangle\!\rangle}} 
\newcommand{\eav}[1]{\ensuremath{\left\langle #1 \right\rangle} } 
\newcommand{\prob}{\ensuremath{\mathbb P}}

\newcommand{\dissfun}{\ensuremath{\tilde{\Omega}}}
\newcommand{\invo}{\ensuremath{{\mathcal{I}}}}

\newcommand{\shei}[1]{\ensuremath{r}_{#1}}
\newcommand{\swid}[1]{\ensuremath{\hat{r}}_{#1}}

\newcommand{\Revision}{601c74c}

\begin{document}
\date{\today --- \now --- Revision:~\Revision}

\title[]{A microscopic perspective on stochastic thermodynamics}

\author{Bernhard Altaner}
\author{J\"urgen Vollmer}
\affiliation{Max Planck Institute for Dynamics and Self-Organization (MPI DS), Am Fassberg 17, 37077 G\"{o}ttingen, Germany}
\affiliation{Faculty of Physics, Georg-August University G\"{o}ttingen, 37077 G\"{o}ttingen, Germany}



\begin{abstract}
 We consider stochastic thermodynamics as a theory of statistical inference for experimentally observed fluctuating time-series.
 To that end, we introduce a general framework for quantifying the knowledge about the dynamical state of the system on two scales:
 a fine-grained or microscopic, deterministic and a coarse-grained or mesoscopic, stochastic level of description.
 For a generic model dynamics, we show how the mathematical expressions for fluctuating entropy changes used in Markovian stochastic thermodynamics emerge naturally.
 Our ideas are conceptional approaches towards 
 (i)~connecting entropy production and its fluctuation relations in deterministic and stochastic systems and
 (ii)~providing a complementary information-theoretic picture to notions of entropy and entropy production in stochastic thermodynamics.
\end{abstract}


\maketitle

\section{Introduction}
\label{sec:intro}
Complex many-body systems exhibit structure and dynamical phenomena on multiple scales.
Reductionism assumes that dynamics at a given scale are a consequence of the dynamics of some (more) fundamental entities at a smaller scale.
In spite of reductionism there are coarse, so-called effective, dynamical theories emerging on multiple scales~\cite{Anderson1972}.
For instance, in order to describe the dynamics of a cup of water we use hydrodynamics --- and not the equations of motion for $10^{24}$ water molecules.
Besides being more efficient in terms of calculations, using an effective description is often the only possibility to make arrive at any dynamical prediction about complex many-particle systems.
In general, we simply do not have access to the information about the exact microscopic configuration.
While we are used to work with effective theories, their existence is not a trivial fact:
if the temporal evolution on the fundamental level, and thus, by reductionism, also its effective evolution on a coarse level is determined by the initial microscopic state~\cite{Penrose1970}, how can we have effective theories?
Why is the information about the exact initial microscopic state redundant with respect to the coarse-grained evolution?

The theory of complex systems approaches this puzzle using concepts from statistical mechanics and the theory of deterministic dynamical systems~\cite{Crutchfield2012}.
We understand statistical mechanics as a theory about the description of physical systems on multiple scales.
Thus, we follow Jaynes' interpretation~\cite{Jaynes1957,Jaynes1965} and regard thermodynamic entropy as the same concept as entropy in information theory.
According to Jaynes, the best, because least-biased, guess about the probabilistic state $p_x$ maximizes the entropy functional 
$ \entfun[\set{p_i}] = - \sum_i p_i \ln p_i$.
In this so-called \emph{MaxEnt} principle, the maximization is performed with respect to macroscopic constraints, which are formalized by specifying the known value of macroscopic averages.

While today Jaynes' view is commonly accepted as a valid approach to equilibrium statistical mechanics, it is not clear if and how it extends to dynamical, \ie{} nonequilibrium, situations.
The field known today as stochastic thermodynamics provides a thermodynamic interpretation of time-series generated by stochastic processes that model small systems in nonequilibrium environments~\cite{Seifert2012}.
Taking the information-theoretic nature of entropy seriously has recently led to the development of ``information thermodynamics'' as a subfield of stochastic thermodynamics \cite{Sagawa+Ueda2010,Sagawa2012,Mandal+Jarzynski2012,Horowitz_etal2013}.
Using this framework, scientists have succeeded in a formalization and experimental demonstration of the famous thought experiments by Maxwell, Szilard and Landauer~\cite{Landauer1961,Bennett2003} regarding the thermodynamic aspects of information processing~\cite{Toyabe_etal2010,Berut_etal2012}.
These results strengthen the view of thermodynamic dissipation as information that is dynamically written to unobservable degrees of freedom.

In the present work, we complement this progress with a microscopic, deterministic perspective on stochastic thermodynamics.
We take the perspective that a system and its medium are identified by observable and unobservable degrees of freedom, respectively.
Assuming that observable stochastic time-series are consistent with a deterministic microscopic evolution, we introduce two non-stationary phase space ensembles.
The fine-grained ensemble contains information about the history of the evolution of a system on the microscopic scale.
The coarse-grained ensemble is obtained by a \textit{MaxEnt} principle and represents our best guess of the unobservable microscopic state, if we only know the frequencies of observable measurement outcomes.
The evolution of the relative entropy~\cite{Kullback.Leibler1951} of these two descriptions quantifies the information lost to hidden degrees of freedom.
As our main results, we show that (i) the relative entropy is consistent with the notion of dissipation in thermostated non-equilibrium molecular dynamics~\cite{Gallavotti.Cohen1995,Evans.Morriss2008,Jepps.Rondoni2010},  and (ii) that for generic class of models yielding Markovian time-series~\cite{Gaspard1992,Vollmer2002}, the fluctuating notions of entropy from stochastic thermodynamics emerge.
We further show how deterministic~\cite{Evans.Searles2002} and stochastic~\cite{Seifert2012} fluctuation relations are unified and comment on the dissipation rate as a consistency criterion between dynamical theories. 

This work is structured as follows.
In Section~\ref{sec:two-levels} we establish our information-theoretic framework and show its consistency with physical notions of entropy.
In Section~\ref{sec:nmbm} we introduce network multibaker maps (NMBM) as model system and show how the emergence of the expressions known from Markovian stochastic thermodynamics.
Section~\ref{sec:discussion} we discuss the relevance of our result in the context of the common deterministic and stochastic models of physical dynamics.

\begin{figure*}[t]
  \centering
  \includegraphics[width=\textwidth]{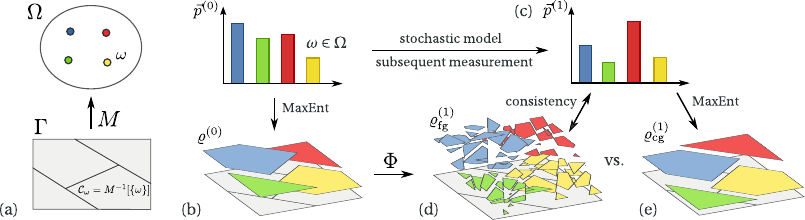}
  \caption{(a) The measurement observable $\partmap$ partitions phase space into disjoint cells $\cell_\omega$ (bottom) indexed by $\omega\in\ospace$ (top). 
    (b) An initial phase space ensemble $\pden\tind{0}$ is obtained from the coarse-grained ensemble $\vec{p}\tind{0}$ by a maximum entropy (\textit{MaxEnt}) principle applied to each cell $\cell_\omega$.
    (c) An iteration of a coarse-grained model or subsequent measurements on a large number of systems yield an updated coarse-grained ensemble $\vec{p}\tind{1}$.
    (d) The microscopic dynamics $\dyn$ propagates $\pden^{0}$ to the fine-grained ensemble $\fgden\tind{1}$.
    It shows an intricate structure that carries information about $\Phi$ and $\pden\tind{0}$.
    (e) On the coarse-grained level, one is ignorant of the microscopic dynamics.
    \textit{MaxEnt} yields the updated coarse-grained ensemble $\cgden\tind{1}$.
}
  \label{fig:two-levels}
\end{figure*}

\section{A dynamical information-theoretic framework for complex systems}
\label{sec:two-levels}

\subsection{Microscopic deterministic and mesoscopic stochastic dynamics}

Consider a complex physical system.
By \emph{complex} we mean that the system exhibits structure on different hierarchical levels.
As examples, think of a complex fluid (like colloids immersed in a solvent) or a biological macromolecule.
A mathematical model of a complex system is formulated at a certain scale, depending on the phenomena it intends to capture.
A well-known model are Hamilton's equations of motion.
Using Hamiltonian dynamics to describe the dynamics of elementary constituents (like atoms, molecules etc.) is often considered as an approach based on first principles.
Then, however, Hamiltonian dynamics is limited to the description of closed, isolated systems. 

Hamiltonian dynamics require us to treat each individual degree of freedom of the environment of the actual (sub-)system of interest.
For the examples above, this requires the solution of the equations of motion for each molecule of the solvent or the cell cytosol --- in spite of the fact that their role in the system is only that of thermodynamic bath, \ie{} a reservoir for heat, momentum, other particles \textit{etc}.
In molecular dynamics simulation, one  uses so-called thermostated equations of motion which use artificial degrees of freedom as an effective description of the environment~\cite{Evans.Morriss2008,Hoover1983}.
Henceforth, we treat Hamiltonian and thermostated deterministic equations of motion on the same footing, and refer to them as the \emph{microscopic} dynamics on a (usually high-dimensional) phase space $\pspace$.

Equations of motion in the form of coupled differential equations specify an evolution rule, which is continuous in time.
Using stroboscopic maps or Poincar\'e sections, one may arrive at a discrete evolution rule specified by an iterated map $\dyn\colon\pspace\to\pspace$ from phase space $\pspace$ onto itself.
Using discretized time steps makes sense from an experimental point of view, because the temporal resolution of any observation is finite.
For the rest of this work we require that $\dyn$ is sufficiently nice such that the Jacobian determinant $\ja(x):= \abs{\det{(\Df\dyn)(x)}}$ exists and is non-zero for almost all microstates $x\in\pspace$.

Due to a finite spatial resolution, the microstates $x\in\space$ cannot be observed directly in experiments.
In most cases, one is not even interested in the exact microstate because.
For instance, one is usually not interested in all rotational degrees of freedom of all amino acids in a large protein.
Instead, one is interested in collective degrees of freedom like its geometric shape, which ultimately determines its function.
Unlike microscopic states, such coarse-grained \emph{mesoscopic} states can be measured in modern experiments.
Formally, a measurement observable $\partmap\colon x\mapsto \omega$ assigns an observable mesoscopic state denoted by an integer $\omega \in \ospace$ to each microstate $x$.
Here we consider the case where $\ospace$ is finite and enumerate the $N$ distinct measurement results $\omega\in\ospace:= \set{1,2,\cdots,N}$ by positive integers.
Thus, $\partmap$ induces a finite disjoint partition of $\pspace = \bigsqcup_{\omega=1}^N \cell_\omega$ into phase space cells $\cell_\omega:=\partmap^{-1}[\set{\omega}]$, where $\partmap^{-1}$ denotes the pre-image operator, \cf~Fig.~\ref{fig:two-levels}(a).
The time-series $\traj\omega(x_0):=\ifam{\partmap\dyn^k x_0}_{k\in\naturals}$ is thus a coarsened description of the microscopic orbit $\traj x(x_0) :=\ifam{\dyn^kx_0}_{k\in\naturals}$ of an initial microstate $x_0$.
Another consequence of a finite experimental resolution is that we do not know the initial microstate $x_0$ of a system.
In statistical physics, initial conditions are specified by a probability density $\pden_0\colon\pspace\to\reals$.
Then, the mesoscopic time-series $\traj \omega(x_0)$ becomes a sequence of random variables, \ie~a stochastic process defined by $\dyn$, $\partmap$ and $\pden\tind{0}$.

Stochastic processes do not necessarily need an underlying microscopic process for their definition.
They are equally well-defined by specifying a consistent probability $\prob[\traj\omega\tind{\tau}]$ for all time series $\traj{\omega}\tind{\tau} = (\omega_0,\omega_1,\dots,\omega_\tau)$ of finite run length $\tau>0$ (via the Kolmogorov extension theorem, see \eg{} Ref.~\cite{Tao2011}).
The most commonly used stochastic process are memoryless, so-called Markov processes, and can be defined in an easy way:
for any two states $\omega,\omega' \in \ospace$, one specifies the conditional probability $0\leq\tprob{\omega}{\omega'}\leq1$ of finding the system in state $\omega'$ at time $t+1$ if the system was in state $\omega$ at time $t$.
Markovian time-series probabilities thus obey
\begin{align}
  \prob\tsub{M}[\traj\omega\tind{\tau}]=  p\tind{0}_{\omega_0} \prod_{k=1}^{\tau} \tprob{\omega_{k-1}}{\omega_k},
  \label{eq:markov-trajectory}
\end{align}
where $p\tind{0}_{\omega_0}$ is a mesoscopic initial condition.
Marginalization shows that the mesoscopic ensemble $\vec{p}\tind{\tau}=(p_\omega\tind{\tau})_\omega$ at time $\tau$ evolves according to the discrete-time master equation
\begin{align}
  p_{\omega'}\tind{t+1}=\sum_{\omega} \tprob{\omega}{\omega'} p_{\omega}\tind{t}.
  \label{eq:master-equation}
\end{align}

%
%
%
%

\subsection{Consistency, information and dynamical entropies}
Let us now return to the microscopic picture.
If our observations are limited to mesoscopic outcomes, $\omega \in \ospace$, our information about the microscopic state of a system is limited.
Besides some thermodynamic or macroscopic information about the set-up, an experimenter may only specify the frequency $p\tind{0}_\omega$ with which she succeeds in preparing the system in a mesoscopic state $\omega$.
In accordance with Jaynes' view, the microscopic initial ensemble should be the least biased distribution that is compatible with this information~\cite{Jaynes1957,Jaynes2003}.
In order to find this distribution we apply the \textit{MaxEnt} principle.
For a continuous phase space it amounts to maximizing the so-called differential entropy
\begin{align}
  \entfun\left[ \pden \right]:=-\int_\pspace\pden\ln\pden \df x
  \label{eq:entfun}
\end{align}
under constraints.
Knowing that the system is in state $\omega$ constrains the probability density to be zero for each microstate $x\neq\cell_\omega$ that is not in phase space cell $\cell_\omega$.
Henceforth,  $\priorden_\omega$ denotes a \emph{localized} \textit{MaxEnt} distribution:
It is supported on $\cell_\omega$ only and maximizes Eq.~\eqref{eq:entfun} under any constraints that formalize additional knowledge about the mesoscopic state $\omega$.
If this information is a thermodynamic statement about the environment of the system, such a localized \textit{MaxEnt} density expresses the assumption of ``local equilibrium'':
the  microstates in each mesoscopic cell are distributed according to constrained equilibrium distributions, \cf also the Appendix of Ref.~\cite{Seifert2011}.
If no additional information is present, \textit{MaxEnt} yields a flat distribution with $\priorden_\omega(x\in\cell_\omega) = \abs{\cell_\omega}^{-1}$, where $\abs{\cell_\omega}$ is the volume (Lebesgue measure) of cell $\cell_\omega$, \cf~Fig.~\ref{fig:two-levels}(b).

Combining the statistical information in the mesoscopic ensemble with the localized \textit{MaxEnt} distributions yields the consistent microscopic initial condition
\begin{align}
  \pden\tind{0}(x) :=  \sum_\omega \left[ p\tind{0}_\omega \priorden_\omega(x)\right].  \label{eq:InitialDensity}
\end{align}
It evolves according to the Frobenius--Perron theorem~\cite{Hopf1948,Ruelle1999} and can be written as
\begin{align}
  \fgden\tind{\tau}(x) := \frac{\pden\tind{0}\left( \dyn^{-\tau} (x) \right)}{\ja\tind{\tau}(\dyn^{-\tau}(x))},
  \label{eq:density-evolution}
\end{align}
where $\ja\tind{\tau}(x) := \prod_{k=1}^\tau \ja(\dyn^{k}(x))$.
Over the course of time, the dynamics introduce microscopic correlations between different parts of phase space, which lead to a more and more complicated structure of the density, \cf{} Fig.~\ref{fig:two-levels}(d).

Any physical model, deterministic or stochastic, is only valid if its predictions reflect experimentally observed results.
At time $\tau$ after preparation, an experimenter measures the mesoscopic distribution $\vec{p}\tind{\tau}$.
In the microscopic picture of a deterministic dynamics, this probability is obtained as an integral of $\fgden\tind{\tau}$ over $\cell_\omega$.
In the mesoscopic picture we find it by marginalizing the time-series probability $\prob[\traj\omega\tind{\tau}]$ on its final state.
Henceforth, we assume mutual consistency between the deterministic and the stochastic model: 
\begin{align}
  \int_{\cell_\omega}\fgden\tind{\tau}\df x \stackrel{!}{=}p\tind{\tau}_\omega \stackrel{!}{=} \sum_{\traj\omega\tind{\tau-1}} \prob[(\traj\omega\tind{\tau-1},\omega)].
  \label{eq:consistency}
\end{align}

Our goal is to quantify and relate the information contained in the fine-grained ensemble $\fgden\tind{\tau}$ with the information in the mesoscopic ensemble $\vec{p}\tind{\tau}$ at finite times $\tau>0$.
To put both on an equal statistical footing, we apply the \textit{MaxEnt} principle to $\vec{p}\tind{\tau}$ and obtain the coarse-grained density
\begin{align}
  \cgden\tind{t}(x) :=  \sum_\omega\left[ p\tind{t}_\omega \priorden_\omega(x)\right],
  \label{eq:cgden}
\end{align}
as the least biased ensemble inferred from the mesoscopic observations at times $\tau>0$.
While equality $\cgden\tind{0}=\fgden\tind{0}\equiv\pden\tind{0}$ holds initially, we have $\cgden\tind{t}\neq\fgden\tind{t}$ for $t>0$, \cf{} Fig.~\ref{fig:two-levels}(d,e).

The uncertainty of a microstate in both ensembles is quantified by their differential entropies \eqref{eq:entfun}.
Hence, we define the dynamical, \ie{} time-dependent, coarse- and fine-grained entropies
\begin{align*}
 \cgent\tind{t} := \entfun\left[ \cgden\tind{t} \right],\quad
 \fgent\tind{t} := \entfun\left[ \fgden\tind{t} \right], 
\end{align*}
 respectively.
Further, we compare the uncertainty in both descriptions relative to one another by means of the following definitions.
The cross entropy 
\begin{align}
  \crossentfun[\pden\Vert\pden'] := -\int \pden \ln \pden'\, \df x
  \label{eq:crossentfun}
\end{align}
measures the average uncertainty of events drawn from an (unknown) distribution $\pden$ that is approximated or modelled by another distribution $\pden'$.
By definition, the cross-entropy $\crossentfun[\pden\Vert\pden']$ is larger than the entropy $\entfun[\pden]$ by the (positive) quantity
\begin{align}
  \dklfun\left[ \pden\Vert\pden '\right]:=\int_\pspace\pden\ln\frac{\pden}{\pden'}\df x = \crossentfun[\pden\Vert\pden'] - \entfun[[\pden],
  \label{eq:dklfun}
\end{align}
known as the (directed) Kullback--Leibler divergence of $\pden'$ from $\pden$~\cite{Kullback.Leibler1951}.
It is also called the relative entropy of $\pden$ with respect to $\pden'$.
By definition, it vanishes if and only if $\pden =\pden'$ and is positive otherwise.

Consequently, we define the relative entropy of the fine-grained with respect to the coarse-grained phase-space density as 
\[\relent\tind{t}:=\dklfun\left[ \fgden\tind{t}\Vert\cgden\tind{t} \right]. \]
It is the information lost when, by assuming $\cgent\tind{t}$, we forget about the dynamic correlations of initial conditions introduced in $\fgden\tind{t}$, over the course of time.
Note that in contrast to $\cgent$ and $\fgent$, $\relent$ is invariant under coordinate transformations or the change of reference measure~\cite{Touchette2009,Polettini2012}.

\subsection{Fluctuating entropies as random variables}

The entropies $\cgent$, $\fgent$ and $\relent$ (defined by the expressions \eqref{eq:entfun} and \eqref{eq:dklfun}) can be interpreted as phase-space (ensemble) averages $\eav{\entrv} := \int_\pspace \pden\,\entrv\,\df{x}$ of a random variable $\entrv(x)$.
The relation between mesoscopic time-series $\traj \omega$ and phase-space averages is made as follows:
Let  $\traj{\omega}\tind{\tau} = (\omega_0,\omega_1,\dots,\omega_\tau)$ be a fixed time series of finite run length $\tau>0$.
Then, all microscopic initial conditions $x_0$ starting in the set
\begin{align*}
  \cell\!\left[ \traj{\omega}\tind{\tau} \right]:= \bigcap_{k=0}^\tau \dyn^{-k}\!\left[ \cell_{\omega_k} \right] = \set{x_0\,\middle\vert\,\forall 0\leq k\leq \tau\colon M\dyn^kx_0 = \omega_k}
\end{align*}
yield microscopic orbits $\traj{x}\tind{\tau}(x_0) = (\dyn^kx_0)_{0\leq k\leq \tau} $  which are mesoscopically indistinguishable.
By definition, their observed time-series obey 
\begin{align*}
  \traj\omega\tind{\tau} =  \traj{\omega}\tind{\tau}(x_0) := (\partmap \dyn^kx_0)_{0\leq k\leq \tau} . 
\end{align*}
For each $\tau$, the sets $\cell\!\left[ \traj{\omega}\tind{\tau} \right]$ obtained from all possible time-series $\traj\omega\tind{\tau}$  form a disjoint partition of phase space.
Hence, the average over any phase-space function $s\tind{\tau}(x)=s[\traj{\omega}\tind{\tau}(x)]$ that only depends on the microstate $x$ via its mesoscopic time series $\traj{\omega}\tind{\tau}(x)$  can be written as
\begin{align}
  \int_\pspace\pden\tind{0}(x)s\tind{\tau}(x)\, \df x = \sum_{\traj \omega\tind{\tau}} \prob_\dyn\left[ \traj{\omega}\tind{\tau} \right] s[\traj{\omega}\tind{\tau}]=:\trav{s}\tind{\tau},
  \label{eq:trav}
\end{align}
where the probability of a time-series is $\prob_\dyn\left[ \traj{\omega}\tind{\tau} \right]:=\int_{ \cell\!\left[ \traj{\omega}\tind{\tau} \right]}\pden\tind{0}\,\df x$.

The probabilities $\prob_\dyn$ define the \emph{stochastic process generated by the microscopic dynamics} $\Phi$.
By definition, this stochastic process it is consistent with the microscopic dynamics, \ie{} it satisfies Eq.~\eqref{eq:consistency}.
Note that besides the stochastic process $\prob_\dyn$ there may be other stochastic processes (\eg{} Markovian processes obeying Eq.~\eqref{eq:markov-trajectory}), for which Eq.~\eqref{eq:consistency} holds.
In general, the stochastic processes generated by the microscopic dynamics $\prob_\dyn$ is \emph{not} strictly Markovian, though the Markov property might be a valid approximation.
For the rest of this work we will always refer to the time-series probabilities $\prob_\dyn$ generated by the dynamics $\dyn$ from an initial measure $\pden\tind{0}$ and thus simply write $\prob$ instead of $\prob_\dyn$.

\begin{figure*}[t]
  \centering
  \includegraphics[width=\textwidth]{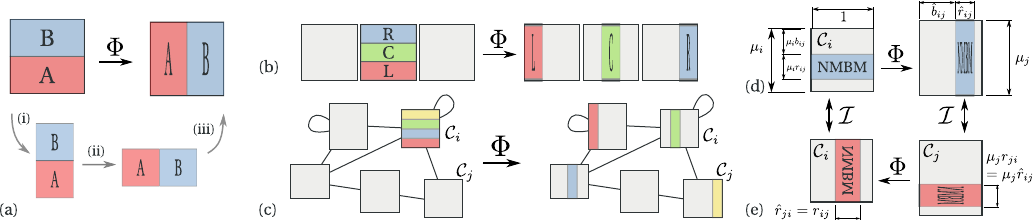}
  \caption{The classical baker map (a) maps the unit square onto itself.
    It can be composed of three steps:
    (i) a linear contraction along the horizontal coordinate,
    (ii) an affine displacement of the two strips $A$ and $B$,
    (iii) a linear expansion along the vertical coordinate.
    A multibaker map (b) additionally displaces the strips along a linear chain of identical baker cells.
    Network multibaker maps (c) are the natural generalization to arbitrary topologies.
    (d) For a reversible network multibaker map, the relative widths of horizontal and vertical strips obey $\shei{ij} = \swid{ji}$.
    Then, the inverse dynamics obeys $\dyn^{-1} = \invo \dyn \invo$, where $\invo$ is a measure-preserving involution that factors to the cells $\cell_j$.
    Geometrically, the involution involves a vertical rescaling of the cell to the unit square followed by a reflection along the second diagonal followed by the reverse vertical rescaling.
  }
  \label{fig:bakermap}
\end{figure*}
In the following, we consider four  time-series dependent functionals.
Three of them depend on the time-series only through the state  of the system at a time $\tau$: 
\begin{align}
  \obsentrv\tind{\tau}\left[  \traj{\omega}\tind{\tau}  \right] &:= -\ln p\tind{\tau}_{\omega_\tau}  \label{eq:obsentrv}\\
  \priorentrv\left[  \traj{\omega}\tind{\tau}  \right]  &:= -\int_{\cell_{\omega_\tau}}\priorden_{\omega_\tau} \ln \priorden_{\omega_\tau}\,\df x  \label{eq:priorentrv}\\
  \crossentrv\tind{\tau}\left[  \traj{\omega}\tind{\tau}  \right]&:=-\int_{\cell_{\omega_\tau}}\left({\fgden\tind{\tau}}/{p\tind{\tau}_{\omega_\tau}}\right) \ln \priorden_{\omega_\tau}\,\df x  \label{eq:crossentrv}
%
\end{align}
The first quantity, $\obsentrv\tind{\tau}$ is the self-information or reduced uncertainty associated with finding the system in the mesoscopic state $\omega_\tau$ at time $\tau$, if the coarse-grained ensemble $\vec{p}\tind{\tau}$ is known.
The second quantity $\priorentrv$ is the entropy of the distribution $\priorden_{\omega_\tau}$, which was obtained by the \textit{MaxEnt} principle introduced above.
Note that $\priorentrv$ does not explicitly depend on time.
The quantity $\crossentrv\tind{\tau}$ is the cross-entropy $\crossentfun$ obtained if the fine-grained distribution $\fgden\tind{\tau}$ marginalized to $\cell_{\omega_\tau}$ is approximated by the localized \textit{MaxEnt} distribution $\priorden_{\omega_\tau}$.

In contrast to the first three quantities, the fourth quantity depends on the complete history of a time series $\traj{\omega}\tind{\tau}$.
It is the averaged phase space contraction factor experienced by the mesoscopically co-moving microscopic orbits $x\in\cell\left[ \traj{\omega}\tind{\tau} \right]$: \begin{align}
  \contentrv\tind{\tau}\left[  \traj{\omega}\tind{\tau}  \right] := -\int_{\cell\left[ \traj{\omega}\tind{\tau} \right]}(\pden\tind{0}/\prob[ \traj{\omega}\tind{\tau}   ]) \ln \ja\tind{\tau} \,\df x.
  \label{eq:contentrv}
\end{align}
It quantifies the irreversibility of the microscopic orbits producing certain mesoscopic observations.
As such, it is related to the notion of dissipation in thermostated dynamics \cite{Evans.Searles2002,Jepps.Rondoni2010}, which we discuss in more detail below. 

A brief calculation~\cite{Altaner2014} 
shows that the fundamental fluctuating entropies \eqref{eq:obsentrv}--\eqref{eq:contentrv} can be used to express the information-theoretic entropies $\cgent\tind{\tau} $, $\fgent\tind{\tau} $ and $\relent\tind{\tau} $ as time-series averages:
\begin{align}
  \cgent\tind{\tau}&:= \trav{\obsentrv + \priorentrv}\tind{\tau},\label{eq:cgentrv}\\
  \fgent\tind{\tau}&:= -\trav{\contentrv}\tind{\tau} + \entfun[\pden\tind{0}], \label{eq:fgentrv}\\
  \relent\tind{\tau}&:= \trav{\obsentrv + \crossentrv + \contentrv}\tind{\tau} -\entfun[\pden\tind{0}] .
  \label{eq:relentrv}
\end{align}

%

\subsection{Entropy functionals for the system and the medium}

So far, we have compared the information contained in two phase-space distributions which formalize our knowledge about consistent microscopic and mesoscopic dynamics.
Next, we discuss the entropy of the system and its surrounding medium from an operational point of view.

By operational we mean that the system and its medium are distinguished by the measurement procedure, \cf{} Refs.~\cite{Altaner2014,Penrose1970,Ruelle1999,Maes2004}.
In this distinction the system contains the degrees of freedom $\omega \in \ospace $ that we can (or choose to) observe in an experiment.
The entropy of the medium characterizes the dynamic uncertainty about the remaining, unobserved, degrees of freedom.

On the level of a single time-series $\traj\omega\tind{\tau}$ of length $\tau$, the definition of the system's entropy $\sysentrv$ is easy.
It is the self-information $\obsentrv=-\ln p\tind{\tau}_{\omega_\tau}$ of sampling the final state $\omega_\tau$ from the mesoscopic ensemble $\vec p \tind{\tau}$ at time $\tau$.
Hence, 
\begin{subequations}
  \begin{align}
    \sysentrv\tind{\tau}  &:= \obsentrv\tind{\tau}
    \label{eq:derived-sysentrv}
  \end{align}
The entropy associated to the medium is more involved and contains two terms.
One term reflects the uncertainty we have regarding the microstates at a given time $\tau$ when we assume local equilibrium, \ie{} the existence of a constrained \textit{MaxEnt} distribution $\priorden_\omega$ on cell $\cell_\omega$, \cf{} also Ref.~\cite{Seifert2011}.
It is quantified by the cross entropy $\crossentrv\tind{\tau}$ discussed above.
The other part regards the correlations in the unobservable microscopic degrees of freedom that are introduced by the microscopic dynamics $\dyn$.
They are quantified by the phase-space contraction factor $\ln \ja(x)$  averaged over all microscopic orbits compatible with the time-series $\traj \omega\tind{\tau}$, \ie{} by $\contentrv\tind{\tau}$, Eq.~\eqref{eq:contentrv}.
Consequently:
  \begin{align}
    \medentrv\tind{\tau}  &:= \crossentrv\tind{\tau}  +\contentrv\tind{\tau}. 
  \end{align}
  \label{eq:derived-entrv}
\end{subequations}
The total entropy is the sum of both contributions:
\begin{align}
  \totentrv\tind{\tau} := \sysentrv\tind{\tau}+\medentrv\tind{\tau} 
 \equiv \obsentrv\tind{\tau}+ \crossentrv\tind{\tau} +\contentrv\tind{\tau} 
  \label{eq:totentrv}
\end{align}

We will analyse the significance of these definitions in the context of thermostated dynamics in more detail below.
For now, we are satisfied with the following consistency check regarding the average total entropy $\totent\tind{\tau} := \trav{\totentrv}\tind{\tau}$ at time $\tau$. 
Eq.~\eqref{eq:relentrv} implies that it emerges as the sum of the relative entropy plus the initial entropy, \ie{} $\totent\tind{\tau} = \relent\tind{\tau} + \entfun[\pden\tind{0}]$.
It is consistent with the second law, because the change in total entropy $\Delta\totent\tind{\tau} := \totent\tind{\tau}-\totent\tind{0} = \relent\tind{\tau}$ in the interval $[0,\tau]$ is a Kullback--Leibler divergence and thus always positive.
Moreover, the total entropy agrees with the fine-grained entropy at time $\tau=0$.
Yet, unlike the fine-grained entropy $\fgent\tind{\tau}$, it is not constant for Hamiltonian-like dynamics but increases until a coarse-grained steady-state is reached.
It thus accounts both for transient observable irreversibility (while the mesoscopic distribution relaxation to its coarse-grained steady state) as well as for the intrinsic microscopic irreversibility of the dynamics.

\section{A microscopic model for Markovian stochastic thermodynamics}
\label{sec:nmbm}

So far we were mainly concerned with definitions and showed their mutual consistency.
Next, we introduce a deterministic microscopic model dynamics in order to study the fundamental fluctuating entropies \eqref{eq:obsentrv}--\eqref{eq:contentrv} explicitly.
The goal is to relate the the microscopically motivated quantities \eqref{eq:derived-entrv} to the expressions used in Markovian stochastic thermodynamics~\cite{Seifert2012}.
Recall that we introduced Markov chains as examples of stochastic processes above.
However, the discussion so far did not assume Markovian trajectory probabilities Eq.~\eqref{eq:markov-trajectory}. 

In this section, we introduce a generic and versatile class of two-dimensional hyperbolic systems, where all quantities introduced in Sec.~\ref{sec:two-levels} can be explicitly calculated.
In particular, our model dynamics give rise to a Markovian coarse-grained evolution.
Then, we use the Master equation \eqref{eq:master-equation} to calculate the system entropy $\sysentrv=\obsentrv$ for any $\tau\geq0$.
Moreover, our microscopic dynamics are analytically tractable, such that we can calculate the other quantities \eqref{eq:priorentrv}--\eqref{eq:contentrv} and their time-series averages as well.
From that we obtain the medium entropy $\medentrv[\traj\omega\tind{\tau}]$ associated to an individual time-series.
We will see that our results yield a microscopic perspective on Markovian stochastic thermodynamics.

\subsection{Reversible network multibaker maps}
Our model is an extension of baker maps introduced by Hopf in the context of early ergodic theory \cite{Hopf1948}, see Fig.~\ref{fig:bakermap}(a).
Gaspard and co-workers studied area-preserving multibaker maps --- that is, a linear chain of coupled baker maps \cite{Gaspard1992,Gilbert.etal2000}, see Fig.~\ref{fig:bakermap}(b) --- as examples for the mathematicians' dynamical-systems approach to statistical mechanics, \cf~Refs.~\cite{Sinai1968,Bowen.Ruelle1975,Ruelle2004}.
Vollmer and co-workers discussed more general, reversible multibaker maps and their connection to reversible thermostats \cite{Vollmer.etal1997,Breymann.etal1998,Vollmer2002}.
Being two-dimensional, (multi)baker maps provide a generic model for hyperbolic dynamics \cite{Smale1967}.
Due to their analytical accessibility and their connection with Markovian dynamics, variants of multibaker maps continue to serve as generic tools to investigate fundamental questions of statistical mechanics~\cite{Colangeli.etal2011,Kawaguchi.Nakayama2013}.

Here, we introduce network multibaker maps (NMBM) as the generalization of these dynamics.
The term 'network' indicates that they are generalizations of the common multibaker map from linear chains to the network of states representing arbitrary Markov chains.
The nodes of the network are $N$ rectangular 'baker cells' $\cell_i \simeq [0,1]\times[0,\vcell_i]$.
The  phase space of a NMBM is their disjoint union $\pspace := \bigsqcup_{i=1}^N$ and we understand them as the elements of the partition induced by a measurement observable, \cf~Fig.~\ref{fig:two-levels}(a).
Each cell is considered adjacent to a set of distinct neighboring cells, represented by the edges of a graph, see Fig.~\ref{fig:bakermap}(c).
Note that a cell can be adjacent to itself.
Like in the original baker map, horizontal strips are are mapped to vertical strips in adjacent cells via a linear horizontal contraction, a linear vertical expansion and a displacement.
Thus, NMBM belong to the class of piecewise continuous affine-linear transformations.

For a formal definition of a NMBM, denote by $\shei{ij}$ the relative height of a horizontal strip $\cell_{ij}\subset\cell_{i}$ that is mapped into a vertical strip $\hat\cell_{ij} := \Phi[\cell_{ij}]\subset\cell_j$ of relative width $\swid{ij}$ in an adjacent cell $\cell_j$.
If  two cells $\cell_i$ and $\cell_j$ are not adjacent, we set $\shei{ij} = \swid{ij}=0$.
As each cell is partitioned by its strips, the normalization condition $\sum_j \shei{ij} = \sum_i \swid{ij} = 1$ holds.
Further, let $b_{ij} := \sum_{k<k} \shei{ij}$ and $\hat{b}_{ij} = \sum_{k<i} \swid{kj}$ be vertical and horizontal offsets of the strips in a cell, \cf~Fig.~\ref{fig:bakermap}(d).
Then, a microstate  $x\in\pspace$ can be written as a triple $(x_1,x_2,i)$ where $x_1$ and $x_2$ denote the horizontal and the vertical coordinate, respectively; the last component denotes the cell index.
We have $x \in \cell_{ij}\subset\cell_i$, if and only if $b^i_j<\frac{x_2}{\vcell_i}\leq b^i_{j+1}$.
For such a microstate, the NMBM dynamics is described by the mapping
\begin{align}
  \dyn\colon  (x_1,x_2,i) &\mapsto \left(\hat{b}_{ij} + \swid{ij}x_1,\,\frac{\vcell_j}{ \shei{ij}}\left(\frac{x_2}{\vcell_i} - b_{ij}\right),j \right),  \label{eq:nmbm-analytical}
\end{align}
The measurement observable reads $\partmap\colon (x_1,x_2,i)\mapsto i$. 

Next, note that a NMBM dynamics is consistent with a Markovian coarse-grained evolution.
To that end, we consider the volume element $\cell[\traj\omega\tind{\tau}]$ defined above.
For $\tau=0$, $\cell[(\omega_0)]=\cell_{\omega_0}$ is just the initial baker cell.
For $\tau=1$, $\cell[(\omega_0,\omega_1)] = \cell_{\omega_0\omega_1}$ is the horizontal strip of relative height $\shei{\omega_0\omega_1}$ in $\cell_{\omega_0}$ which is mapped to $\cell_{\omega_1}$.
For $\tau=2$ we have $\cell[(\omega_0,\omega_1,\omega_2)]\equiv \cell_{\omega_0}\cap \dyn^{-1}[\cell_{\omega_2}] \cap \dyn^{-2}[\cell_{\omega_2}] \subset \cell_{\omega_0} \cap \dyn^{-2}[\cell_{\omega_2}]$.
It is immediately clear that $\dyn^{-2}[\cell_{\omega_2}]$ consists of horizontal strips of relative height $\shei{\omega_0,\nu}\shei{\nu,\omega_2}$.
Their number is determined by the distinct mesoscopic time series $(\omega_0,\nu,\omega_2)$ that are possible for points which are mapped from $\omega_0$ to $\omega_2$ by $\dyn^{2}$.
Intersecting this set with $\cell_{\omega_0\omega_1} \equiv \cell_{\omega_0} \cap \dyn^{-1}[\cell_{\omega_1}]$ selects the microstates that generate the time series $(\omega_0,\omega_1,\omega_2)$.
By the same argument, one finds that $\cell[\traj\omega\tind{\tau}]$ is a horizontal strip of $\cell_{\omega_0}$ with relative height $\prod_{k=1}^{\tau} \shei{\omega_{k-1}\omega_k}$ and thus $\abs{\cell[\traj\omega\tind{\tau}]}=\vcell_{\omega_0}\prod_{k=1}^{\tau} \shei{\omega_{k-1}\omega_k}$.
Without additional thermodynamic information about the dynamics, the \textit{MaxEnt} density $\priorden_i=\vcell_i^{-1}$ for $\cell_i$ is uniform and we find 
\begin{align}
  \prob[\traj\omega\tind{\tau}] 
  = \abs{\cell[\traj\omega\tind{\tau}]}\frac{p_{\omega_0}}{\vcell_{\omega_0}}
  = p\tind{0}_{\omega_0} \prod_{k=1}^{\tau} \shei{\omega_{k-1}\omega_k},
  \label{eq:nmbm-markov}
\end{align}
which is consistent with a Markov process with transition probabilities $\shei{ij}$ and thus obeys the Master equation \eqref{eq:master-equation}.

Note that this result does not depend on the choice of $\swid{ij}$ and $\vcell_i$;
for each Markov chain with transition probabilities $\shei{ij}$, there is an infinite number of compatible NMBM.
Henceforth, we are mostly interested in the case of so-called reversible NMBM, which are defined by the symmetry 
\begin{align}
  \swid{ij} = \shei{ji}.
  \label{eq:nmbm-reversibility}
\end{align}
Then, it is easy to check that the map
\begin{align}  
  \invo \colon   (x_1,x_2,i) &\mapsto (1-\vcell_i^{-1}x_2, \vcell_i(1-x_1),i),\,\label{eq:invo} 
\end{align}
is a measure-preserving time-reversal involution obeying
\begin{subequations}
  \begin{align}
    \invo^2&=\mathrm{id},\label{eq:rev-invo}\\
    \dyn^{-1} &= \invo \dyn\invo \label{eq:rev-timerev}.
  \end{align}
Moreover,  $\invo$ is measure-preserving and acts locally on the baker cells $\cell_i$ (see also Fig.~\ref{fig:bakermap}(e)), \ie{}
\begin{align}
  \abs{\invo[A]}&=\abs{A}\text{ for all $A\subset\pspace$}, \label{eq:rev-mp}\\
  \invo[\cell_i] &= \cell_i.\label{eq:rev-local}
\end{align}
  \label{eq:reversibility}
\end{subequations}
In the special case of a linear topology, multibaker maps obeying this property have previously been called 'properly thermostated'~\cite{Vollmer2002}.
We will discuss the role of reversible NMBM maps as generic models for thermostated dynamics below.
Finally, note that microscopic reversibility~\eqref{eq:reversibility} ensures dynamical reversibility of the coarse-grained Markov process, \ie{} the fact that the transition probabilities obey $\swid{ij}>0 \Leftrightarrow = \swid{ji}>0$.

\subsection{Fluctuating entropies for NMBM}
Because of the Markov property \eqref{eq:nmbm-markov}, the self-information  $\obsentrv\tind{\tau}(\omega_\tau) = -\ln p\tind{\tau}_{\omega_\tau}$ of a measurement at time $\tau$ can be computed directly.
Using that $\priorden_{\omega_\tau}$ and (and due to Eq.~\ref{eq:InitialDensity} the initial density $\pden\tind{0}$) is constant on each cell, we find that
\begin{align}
  \crossentrv\tind{\tau}(\omega_\tau) = \priorentrv(\omega_\tau) = -\ln\priorden_{\omega_\tau} \equiv \ln \vcell_{\omega_\tau}.
  \label{eq:nmbm-crossent-priorent}
\end{align}
In order to calculate $\contentrv\tind{\tau}$ we need the value of the Jacobian determinant on $\cell[\traj\omega\tind{\tau}]$.
From the definition of the network multibaker map \eqref{eq:nmbm-analytical}, it directly follows that $\ja(x) = (\vcell_j\swid{ij})/(\vcell_i \shei{ij})\equiv \abs{\hat{\cell}_{ij}}/\abs{\cell_{ij}} $ for all $x\in\cell_{ij}$, expressing the uniform phase-space contraction within a horizontal strip.
Consequently,
\begin{align}
  \ja\tind{\tau} = \prod_{k=1}^\tau \frac{\vcell_{\omega_{k}}\swid{\omega_{k-1}\omega_k}}{\vcell_{\omega_{k-1}} \shei{\omega_{k-1}\omega_k}}\stackrel{\eqref{eq:nmbm-reversibility}}{=} \prod_{k=1}^\tau \frac{\vcell_{\omega_{k}}\shei{\omega_{k}\omega_{k-1}}}{\vcell_{\omega_{k-1}} \shei{\omega_{k-1}\omega_k}}
  \label{eq:nmbm-jac}
\end{align}
is constant for all $x\in\cell[\traj\omega\tind{\tau}]$, and we obtain
\begin{align}
  \contentrv\tind{\tau}[\traj\omega\tind{\tau}]  %
  = -\ln \ja\tind{\tau}
  =\ln\frac{\vcell_{\omega_0}}{\vcell_{\omega_\tau}}+\sum_{k=1}^\tau\ln \frac{ \shei{\omega_{k-1}\omega_k}}{\shei{\omega_{k}\omega_{k-1}}}.
  \label{eq:nmbn-contentrv}
\end{align}
From Eqs.~\eqref{eq:nmbm-crossent-priorent} and \eqref{eq:nmbn-contentrv} we obtain total entropy \eqref{eq:totentrv} as the sum of the contributions from the  system and its surrounding medium \eqref{eq:derived-entrv}:
\begin{align}
  \totentrv\tind{\tau}[\traj\omega\tind{\tau}] &= %
  \underbrace{%
    -\ln p\tind{\tau}_{\omega_\tau}%
  }_{=\sysentrv\tind{\tau}(\omega_\tau)}%
  +%
  \underbrace{%
   \sum_{k=1}^\tau\ln \frac{ \shei{\omega_{k-1}\omega_k}}{\shei{\omega_{k}\omega_{k-1}}} + \ln \vcell_{\omega_0}   %
  }_{\medentrv\tind{\tau}[\traj\omega\tind{\tau}]}.
  \label{eq:nmbm-totentrv}
\end{align}
The change \emph{change} of the fluctuating entropies along the time series $\traj\omega\tind{\tau}$ is defined as $\eprv\tind{\tau}[\traj\omega\tind{\tau}]  := \entrv\tind{\tau}[\traj\omega\tind{\tau}] - \entrv\tind{0}(\omega_0)$.
The last term in Eq.~\eqref{eq:nmbm-totentrv} is a constant contribution.
The change in total entropy production thus reads
\begin{subequations}
\begin{align}
  \toteprv\tind{\tau}[\traj\omega\tind{\tau}] &=\ln\frac{p\tind{0}_{\omega_{0}}}{p\tind{\tau}_{\omega_\tau}} +\sum_{k=1}^\tau \ln\frac{\shei{\omega_{k-1}\omega_{k}}}{\shei{\omega_{k}\omega_{k-1}}}  
\end{align}  
and consists of the individual contributions associated to the system and the medium
\begin{align}
  \syseprv\tind{\tau}[\traj\omega\tind{\tau}] &= \ln\frac{p\tind{0}_{\omega_{0}}}{p\tind{\tau}_{\omega_\tau}},\\
  \medeprv\tind{\tau}[\traj\omega\tind{\tau}] &= \sum_{k=1}^\tau \ln\frac{\shei{\omega_{k-1}\omega_{k}}}{\shei{\omega_{k}\omega_{k-1}}}. \label{eq:nmbm-medeprv}
\end{align}  
  \label{eq:nmbm-final}
\end{subequations}

Remarkably, Eqs.~\eqref{eq:nmbm-final} are exactly the entropic expressions that form the basis of the Stochastic Thermodynamics of Markovian jump processes described by a Master equation~\eqref{eq:master-equation} with transition probabilities $r^i_j$~\cite{Lebowitz.Spohn1999,Maes2004,Seifert2005,Seifert2012}.
However, unlike in previous cases they are not motivated from mesoscopic thermodynamic considerations, but from the fluctuating entropies \eqref{eq:derived-entrv} defined in the context of a deterministic microscopic dynamics.

\section{Discussion}
\label{sec:discussion}
In this section, we discuss the significance of the present result in the context of general deterministic and stochastic dynamics. 
We argue that NMBM serve as a versatile yet analytically tractable model for physical dynamics.
In particular, we point out how deterministic and stochastic fluctuation relations for the dissipation emerge in the context of reversible NMBM models.
We finish the discussion with some general remarks on information, dissipation and consistency in the information-theoretic framework introduced in Sec.~\ref{sec:two-levels}.

\subsection{NMBM dynamics mimic physical  dynamics}
NMBM are  a generic, yet analytically tractable model for physical dynamics.
On the mesoscopic level, NMBM yield Markovian dynamics.
Many of the stochastic processes investigated in physics at least appear Markovian~\cite{vKampen1992}.
Penrose stresses that the Markovian character of a process is crucial for statistical reproducibility of observation:
observable states must not keep a memory of their preparation procedure \cite{Penrose1970}.
Given Penrose's argument it is not surprising that ST is only fully understood for Markovian dynamics. 
In ST, Markovianity is a direct consequence of the assumption of mesoscopic local equilibrium, \ie{} the assumption of a \textit{MaxEnt} principle on the level of the observable mesoscopic states~\cite{Seifert2011,Esposito2012,Esposito.VandenBroeck2010}.
In spite of being low-dimensional, NMBM thus provide a useful tool for investigating the microscopic processes that yield mesoscopic Markovian statistics, \cf{} also Refs.~\cite{Colangeli.etal2011,Kawaguchi.Nakayama2013}.

In the context of MD simulations, one usually considers the equations of motion physical if they are reversible~\cite{Jepps.Rondoni2010}. 
In the narrowest sense, reversibility is defined as the existence of a time-reversal involution $\invo$ that obeys Eqs.~\eqref{eq:rev-invo} and \eqref{eq:rev-timerev}.
In Hamiltonian dynamics, time-reversal is obtained by inverting the momenta of all particles.
In the more general case of thermostated equation of motions, the involution $\invo$ also inverts all anti-symmetric auxiliary variables~\cite{Jepps.Rondoni2010}.
Note that these physical involutions preserve the (Lebesgue or Liouville) measure of phase space, \ie{} they obey Eq.~\eqref{eq:rev-mp}.
In particular, any smooth involution must be measure-preserving as a consequence of the chain rule for the Jacobian determinant.
The locality assumption~\eqref{eq:rev-local} connects the involution formulated for the microscopic dynamics to mesoscopic observables.
For the case of thermostated molecular dynamics, it is valid for all observables with even parity, \eg{} all observables that only depend only on the positions of particles.
Consequently, we consider reversible NMBM obeying Eqs.~\eqref{eq:reversibility} as a valid low-dimensional representation of physical microscopic dynamics.

Moreover, NMBM can be tuned to show features of Hamiltonian, conservative and dissipative systems corresponding to Hamiltonian, thermostated and thermostated-driven systems:\\
(i) Hamiltonian dynamics corresponding to isolated systems uniformly conserve phase space volume.
For discrete dynamics, this condition reads $\log\ja\tind{\tau} = 0$.
Hence, reversible NMBM are uniformly conservative if and only if $\vcell_i\shei{ij} =\vcell_j\shei{ji}$, \cf~ Eq.~\eqref{eq:nmbm-jac}.\\
(ii) Non-uniformly conservative systems are more general and have the property that the phase-space contraction rate $\tau^{-1}\trav{\contentrv}\tind{\tau}$ vanishes for $\tau \to \infty$.
For NMBM,  the phase-space contraction rate~\eqref{eq:nmbn-contentrv} is asymptotically dominated by the rate of change in the medium~\eqref{eq:nmbm-medeprv}.
From the theory of Markov processes \cite{Schnakenberg1976}, we know that this quantity only vanished for detailed balance systems.
The transitions of such systems obey the Kolmogorov cycle criterion, \ie{} it holds that 
\begin{align*}
  \prod_{k=1}^\tau \shei{\omega_{k-1}\omega_{k}} = \prod_{k=1}^\tau \shei{\omega_{k}\omega_{k-1}} 
\end{align*}
for each closed path, \ie~for each $\traj \omega\tind{\tau}$ with $\omega_0 = \omega_\tau$.
Consequently, reversible NMBM are conservative if (and only if) the mesoscopic stochastic dynamics is described by a detailed balance system.
An equivalent definition of an equilibrium system in ST is the vanishing of all cycle affinities~\cite{Schnakenberg1976}.\\
(iii) The generic case obtained for other choices of the number $r^i_j$ is that of a dissipative system.
For such dynamics, the fine-grained entropy diverges to $-\infty$.
The rate of this divergence is interpreted as the dissipation rate~\cite{Evans.Searles2002,Jepps.Rondoni2010}.
The reason for this divergence is that the probability density converges to the fractal distribution of a so-called SRB measure~\cite{Young2002}.
For reversible linear multibaker chains, this fractal structure was studied under periodic boundary conditions~\cite{Vollmer2002}.
The same structure can be obtained also in the case of uniformly conservative multibaker maps in the case of open boundaries or in infinite systems~\cite{Gaspard2005}.
With NMBM, one obtains a model dynamics to investigate these structures on arbitrary topologies of baker cells.

\subsection{Emergence of deterministic and stochastic fluctuation relations}

In the introduction, we mentioned fluctuation relations (FRs) as statistical refinements of the second law of thermodynamics.
In the deterministic case, they are a consequence of the existence of a measure-preserving time-reversal involution~\cite{Gallavotti.Cohen1995,Evans.Searles2002,Wojtkowski2009}.
For stochastic systems, formulating an FR requires the notion of a conjugate process which generates conjugate stochastic trajectories, \ie~conjugate time series \cite{Seifert2012}.
The conjugate process (and thus the conjugate time series average $\trav{\,\cdot\,}\tind{\tau}_*$) is uniquely defined by specifying the probabilities $\prob_*[\Theta\traj\omega\tind{\tau}]$ of conjugate time series.
Then, the  generic fluctuation relation 
\begin{align}
  \frac{\prob[R\tind{\tau}=A]}{\prob[R\tind{\tau}_*=-A]}= \frac{\trav{\delta(R\tind{\tau}-A)}\tind{\tau}}{\trav{\delta(R\tind{\tau}_*+A)}\tind{\tau}_*} =\exp[A]
  \label{eq:fr}
\end{align}
for the quantity
\begin{align}
  R\tind{\tau}[\traj\omega\tind{\tau}] := \ln\frac{\prob[\traj\omega\tind{\tau}]}{\prob_*[\Theta\traj\omega\tind{\tau}]} =:- R\tind{\tau}_*[\Theta\traj\omega\tind{\tau}].
  \label{eq:r}
\end{align}
is a mere consequence of the definitions \cite{Maes2004}, yet with major physical implications \cite{Seifert2005,Seifert2012}.

A key assumption in Markovian ST is dynamical reversibility:
if the transition $\omega \to \omega'$ occurs with non-zero probability, so does the reversed transition $\omega' \to \omega$.
For NMBM dynamical reversibility is a direct consequence of microscopic reversibility, Eqs.~\eqref{eq:reversibility}. 
Hence, it is natural to define the conjugate time series  $\Theta \traj\omega\tind{\tau}=(\omega_{\tau-k})_{0\leq k\leq \tau}$  as the one obtained by reversing the succession of states in a time series $\traj \omega\tind{\tau}$. 
In the following, we consider two choices $\prob_{*,(i)}$ and $\prob_{*,(ii)}$ for the reverse Markovian process, leading to two different log-ratios $R\tind{\tau}_{(i)}$ and $R\tind{\tau}_{(ii)}$, respectively.

(i)~The first one is defined by 
\begin{align*}
  \prob_{*,(i)}[\Theta\traj\omega\tind{\tau}] 
  := p\tind{\tau}_{\omega_\tau}\prod_{k=1}^\tau\shei{\omega_{k}\omega_{k-1}} .
\end{align*}
It is that of Seifert~\cite{Seifert2005} and leads to 
\begin{align}
  R_{(i)}\tind{\tau}[\traj\omega\tind{\tau}] &=\ln\frac{p\tind{0}_{\omega_{0}}}{p\tind{\tau}_{\omega_\tau}} +\sum_{k=1}^\tau \ln\frac{\shei{\omega_{k-1}\omega_{k}}}{\shei{\omega_{k}\omega_{k-1}}}\nonumber \\
  &=  \toteprv\tind{\tau}[\traj\omega\tind{\tau}].
  \label{eq:RSeifert}
\end{align}

Then, the FR \eqref{eq:fr} is the usual notion of the second law in stochastic thermodynamics.

(ii)~The alternative choice 
\begin{align*}
  \prob_{*,(ii)}[\Theta\traj\omega\tind{\tau}] = \prob[\Theta\traj\omega\tind{\tau}]
\end{align*}
 allows for a connection with the Evans--Searles FR regarding the so-called dissipation function \cite{Evans.Searles2002}, which is defined as
\begin{align}
  \dissfun\tind{\tau}(x) := \ln \frac{\pden\tind{0}(x)}{\pden\tind{0}(\dyn(x))} - \Lambda\tind{\tau}(x).
  \label{eq:dissfun}
\end{align}
In the above expression, $\Lambda\tind{\tau}(x)$ denotes the accumulated phase space expansion along the microscopic orbit of $x$ in the interval $[0,\tau]$.
In the current context of iterated maps, we have $\Lambda\tind{\tau}(x) = \ln J\tind{\tau}(x)$.
With the microscopic initial ensemble \eqref{eq:InitialDensity} and expression \eqref{eq:nmbm-jac} for the Jacobian determinant of NMBM, we find
\begin{align}
  \dissfun\tind{\tau}(x) &:= \ln \frac{p\tind{0}_{\omega_0}\vcell_{\omega_\tau}}{p\tind{0}_{\omega_\tau}\vcell_{\omega_0}} - \sum_{k=1}^\tau \ln\frac{\vcell_{\omega_k}\shei{\omega_{k}\omega_{k-1}}}{\vcell_{\omega_{k-1}}\shei{\omega_{k-1}\omega_{k}}}\nonumber\\
  &= \ln \frac{p\tind{0}_{\omega_0}}{p\tind{0}_{\omega_\tau}} + \sum_{k=1}^\tau \ln\frac{\shei{\omega_{k-1}\omega_{k}}}{\shei{\omega_{k}\omega_{k-1}}}\nonumber\\
  &= \ln \frac{\prob[\traj\omega\tind{\tau}(x)]}{\prob_{*,(ii)}[\Theta\traj\omega\tind{\tau}(x)]} \equiv R_{(ii)}\tind{\tau}[\omega\tind{\tau}(x)].
  \label{eq:nmbm-dissfun}
\end{align}
Hence, the fluctuation relation \eqref{eq:fr} applied to $R_{(ii)}\tind{\tau}$ gives the Evans--Searles fluctuation relation.

Asymptotically (\ie{} for $\tau \to \infty$) the expressions $R_{(i)}\tind{\tau}$ and $R_{(ii)}\tind{\tau}$ agree, because they are dominated by the summand $\medeprv\tind{\tau}[\traj\omega\tind{\tau}] \equiv \sum_{k=1}^\tau \ln\frac{\shei{\omega_{k-1}\omega_{k}}}{\shei{\omega_{k}\omega_{k-1}}}$, which typically grows linear in time.
For the case of NMBM, both the deterministic FR and the stochastic FR make the same statement.
The emergence and unification of these FR on both levels of description emphasizes the role of (dynamical) reversibility as a key feature of physically inspired dynamics.
Moreover, it strengthens NMBM as a useful model for addressing conceptional questions regarding the connection between microscopic and the mesoscopic modelling paradigms.

\subsection{Relative entropy and model consistency}
We finish the discussion with some remarks on the general scheme outlined in Sec.~\ref{sec:two-levels}. 
In particular, we argue that the relative entropy, and thus, in our framework, the change in total entropy, fulfils the role of a consistency criterion between dynamical theories formulated on multiple levels.
To make the argument more clear, we revisit the arguments used in the thermodynamic interpretation of thermostated equations of motion and stochastic models.

In the first case, \ie{} in NEMD simulations it is a common assumption that the asymptotic phase-space contraction rate 
\begin{align*}
  \bar\Lambda\tind{\infty} &:=\lim_{\tau\to\infty} \int_\pspace \fgden\tind{\tau} \ja(x)\df x\\
  &=  \lim_{\tau\to\infty} \tau^{-1} \trav{\contentrv}\tind{\tau}
\end{align*}
equals the observable $\sigma\tind{\infty}$, which is interpreted as the dissipation rate in nonequilibrium steady-states~\cite{Gallavotti.Cohen1995,Jepps.Rondoni2010}.
In irreversible thermodynamics \cite{Groot.Mazur1984}, the dissipation rate the scalar product of the vector of macroscopic, physical currents and their conjugate external driving fields.
For thermostated equations of motions under non-equilibrium conditions, average steady-state currents $\set{\eav{J_\alpha}\tind{\infty}}$ and fields $\set{F_\alpha}$ are averages of coarse-grained observables and driving parameters, respectively.
Demanding that the bi-linear form $\sigma :=\sum_\alpha \eav{J_\alpha}\tind{\infty} F_\alpha$ equals the steady-state dissipation rate is thus a consistency requirement between (microscopic) NEMD simulations and (macroscopic) irreversible thermodynamics~\cite{Evans.Morriss2008,Evans.Searles2002}.

In ST, the assumption of constrained local equilibrium distributions for the microstates $x\in\cell_\omega$ forming a coarse-grained state $\omega$ acts as a two-fold consistency criterion between microscopic and mesoscopic dynamics.
The Markovian nature of the dynamics is a consequence of assuming time-independent \textit{MaxEnt} distributions $\priorden_\omega$, which by definition do not carry information about past trajectories.
Additionally, the thermodynamic constraints in the \textit{MaxEnt} principle justifies the definition of the dissipation $\sigma\tind{\infty} = \trav{\medeprv}\tind{\infty}$, \ie{} the entropy changes in the medium~\cite{Seifert2011}.

The information-theoretic framework introduced in Section~\ref{sec:two-levels} unifies these asymptotic consistency using the notion of relative entropy:
\begin{align}
  \sigma\tind{\infty} \stackrel{!}{=} \lim_{\tau\to\infty}\tau^{-1}\relent\tind{\tau}.
  \label{eq:dissipation-relent}
\end{align}
In both the microscopic and the mesoscopic case, the left hand side $\sigma\tind{\infty}$ is interpreted as the steady-state dissipation.
However, the average relative entropy $\relent\tind{\tau} = \trav{\relentrv}\tind{\tau}$ is defined for all finite times.
In particular, Eqs.~\ref{eq:derived-entrv} capture multiple contributions to irreversibility:
(i) the part $\syseprv\tind{\tau}$  which arises from a non-stationary observable probability distribution $\vec{p}\tind{\tau}$ for the observable states as well as
(ii) an additional contribution that originates from the irreversibility of the microscopic dynamics $\dyn$.
Note that the second contributions is present even if the mesoscopic distribution has already relaxed to its steady-state value, and thus the first contribution vanishes.

\section{Conclusion}
In the present work we have introduced a framework for the information-theoretic treatment of complex dynamics on multiple scales.
The \textit{MaxEnt} principle allows us to infer a ``coarse-grained'' phase space density from a dynamically varying observable ensemble.
In addition, we obtained a ``fine-grained'' ensemble from a consistent microscopic deterministic evolution rule on phase space.
In order to quantify the notion of entropy lost to unobservable degrees of freedom, we introduced the relative entropy as the Kullback--Leibler divergence of these two ensembles.
We showed the consistency of the relative entropy with the notion of total entropy production in thermostated dynamics.
Moreover, for a versatile model dynamics yielding Markovian time-series, the fluctuating entropies from stochastic thermodynamics emerge naturally.
In this context, we were thus able to unify deterministic and stochastic fluctuation relations.

Let us restate our main conclusions:
\begin{itemize}
  \item Network multibaker maps provide a useful and generic, yet analytically tractable model for complex deterministic dynamics.
  \item The notion of time-reversal is crucial for fluctuation relations, both in the deterministic and stochastic cases.
  \item The relative entropy between the fine- and coarse-grained ensemble formalizes the dynamical information in hidden degrees of freedom that is inaccessible by coarse-grained measurements.
\end{itemize}
While the first two points provide a conceptional framework for further theoretical studies, the last point links  back to  classical thermodynamics:
after all, the thermodynamic notion of heat is nothing else than energy contained in non-accessible (and thus not exploitable) degrees of freedom.


\acknowledgements

The authors thank Lamberto Rondoni for fruitful discussions during visits to Torino and the KITPC meeting ``Small system nonequilibrium fluctuations, dynamics and stochastics, and anomalous behavior'' held in Beijing in 2013.
Moreover, the authors thank Artur Wachtel for many comments on the various forms of the manuscript.

\bibliography{nmbm_st}


\end{document}